\documentclass[a4paper,11pt]{article}
\usepackage{amsmath}
\usepackage{amssymb}
\usepackage{bbm}
\usepackage{pos}

\newcommand{\ket}[1]{|#1\rangle}
\newcommand{\braOket}[3]{\langle #1 | #2 | #3 \rangle }

\newcommand{\pii}{{\pi_i}}

\newcommand{\calH}{\mathcal{H}}
\newcommand{\calN}{\mathcal{N}}

\newcommand{\lr}[1]{\left( #1 \right)}

\title{Towards Quantum Monte Carlo Simulations at non-zero Baryon and Isospin Density in the Strong Coupling Regime}
 \ShortTitle{Towards QMC Simulations at non-zero Baryon and Isospin Density at Strong Coupling}

\author[a]{Pratitee Pattanaik}
\author*[a]{Wolfgang Unger}

\affiliation[a]{Fakultät für Physik, Universität Bielefeld,\\
  Bielefeld, Germany}

\emailAdd{pratiteep@physik.uni-bielefeld.de}
\emailAdd{wunger@physik.uni-bielefeld.de}

\abstract{
The Hamiltonian formulation of Lattice QCD with staggered fermions
in the strong coupling limit has no sign problem at non-zero baryon density
and allows for Quantum Monte Carlo simulations.

We have extended this formalism to two flavors,
and after a resummation, there is no sign problem
both for non-zero baryon and isospin chemical potential.
We report on recent progress on the implementation
of the Quantum Monte Carlo simulations.
}

\FullConference{%
  The 39th International Symposium on Lattice Field Theory (Lattice2022),\\
  8-13 August, 2022 \\
  Bonn, Germany 
}

\def \Tr {{\rm Tr}}
\def \tr {{\rm tr}}

\def \QMat{\mathcal{M}}
\newcommand{\Qmd}{\QMat^\dagger}

\newcommand{\J}{\mathcal{J}}

\newcommand{\expval}[1]{{\langle #1\rangle }}

\newcommand{\bareT}{\mathcal{T}}
\newcommand{\bareMu}{\mu_{\mathcal{B}}}
\newcommand{\bareI}{\mathcal{\mu_I}}

\newcommand{\meson}{\mathfrak{m}}

\newcommand{\hadron}{\mathfrak{h}}
\newcommand{\spin}{\mathfrak{s}}

\newcommand{\hJ}{\hat{J}}

\newcommand{\SU}{{\rm SU}}

\def \at {{a_{\tau}}}

\def \Nt {{N_{\tau}}}
\def \Nc {{N_{c}}}
\def \Nf {{N_{f}}}

\newcommand{\Hil}{\mathbbm{H}_\hadron}

\newcommand{\nn} {\nonumber\\}

\def \QMat{\mathcal{M}}

\newcommand{\mU}{\pi_U}
\newcommand{\mD}{\pi_D}

\newcommand{\mP}{\pi_+}
\newcommand{\mM}{\pi_-}

\begin{document}
\maketitle

\section{Introduction}

Lattice QCD with staggered fermions in the strong coupling limit has been studied both via Monte Carlo 
\cite{Rossi1984,Karsch1989,Forcrand2010} and mean field theory \cite{Kawamoto1981,Nishida2003} in the last decades.
Whereas the mean field approach is based on a $1/d$ expansion, the formulation suitable for Monte Carlo is a dual representation where the degrees degrees of freedom are color singlets, such as mesons and baryons. It is obtained by integrating out the gauge fields first, after that the Grassmann variables. This formulation has no fermion determinant, but admits a world-line representation. 
In this dual representation, the finite density sign problem is much milder, as the sign only depends on the the geometry of baryonic world-lines. This effective theory of lattice QCD can be very efficiently simulated by the worm algorithm \cite{Forcrand2010}.
It has been extended via the strong coupling expansion to non-zero values of the inverse gauge coupling $\beta=\frac{2\Nc}{g^2}$ \cite{Gagliardi2019}.\\

The main motivation for lattice QCD in the strong coupling regime is that the finite density sign problem is mild enough to study the full
$\mu_B$-$T$ phase diagram. This is still possible if the inverse gauge coupling $\beta$ is not too large \cite{deForcrand2014,Kim2019}.
The drawback of the dual representation is that the sign problem is gradually re-introduced as the lattice gets finer, hence the continuum limit is out of reach. The phase diagram in the strong coupling regime features a critical endpoint at finite quark mass (tricritical in the chiral limit), which for moderate quark masses is located at values much larger than $\mu_{B,c}/T_c>3$ \cite{Kim2016}.
Whether the chiral critical point still exists in the continuum limit is unknown.\\ 

Even though the continuum limit $a\rightarrow 0$ is out of reach in the dual representation, we have studied the continuous Euclidean time limit $a_t\rightarrow 0$, which results in a Quantum Hamiltonian formulation of lattice QCD, where the Euclidean time extend corresponds to the inverse temperature.
Anisotropic lattices with $\xi=a/a_t>1$ are necessary because the spatial lattice spacing $a$ is fixed for fixed $\beta$, and introducing a bare anisotropy $\gamma$ is the only way to continuously vary the temperature $aT=\xi/\Nt$. 
At fixed bare temperature $aT$, the limits $a_t \rightarrow 0$ and $N_t\rightarrow \infty$ are taken simultaneously \cite{deForcrand2017}.
The continuous time limit has many advantages over the formulation on 3+1 dimensional lattice with discrete temporal extent $\Nt$:
\begin{itemize}
 \item The sign problem is completely absent as baryons become static for $a_t\rightarrow 0$.
 \item Ambiguities on the phase boundary present for finite $\Nt$ are remediated.
 \item The dual degrees of freedom can be mapped onto pion occupation numbers.
 \item A quantum Monte Carlo algorithm (continuous time worm algorithm) can be used to directly sample the continuous time partition function.
 \item Continuous time correlation function can be used to determine the hadron masses.
\end{itemize}

The Hamiltonian formulation of lattice QCD has been discussed in detail in \cite{Klegrewe2020} in the strong coupling limit for $\Nf=1$. In contrast to Hamiltonian formulations in the early days of lattice QCD \cite{Kogut1974} this formulation is based on the continuous time limit of the dual representation. 
Whereas in meanfield theory also the extension from $\Nf=1$ flavor of staggered fermions to $\Nf=2$ is straight forward \cite{Bilic1992a,Nishida2003}, the $\Nf=2$ dual formulation is much more involved. As explained in \cite{Unger2021}, the list of color singlet invariants is much larger, and Grassmann integration yields contractions that introduce a severe sign problem also in the mesonic sector. 
However, it was also found that in the continuous time limit, the sign problem is again absent. 
Hence a Quantum Hamiltonian formulation for $\Nf=2$ can be established and can be studied via Quantum Monte Carlo. This allows to study various phenomena that are not present in the $\Nf=1$ formulation:
\begin{itemize}
 \item Simulations at both non-zero baryon and isospin density are possible, hence the phase diagram in the $\mu_B - \mu_I - T$ can be determined
 \item This will also allow to study the relation between pion condensation and the nuclear phase.
 \item Nuclear interactions that are purely entropic for $\Nf=1$ are modified by pion exchange between nucleons.
\end{itemize}

In this proceedings, we will report on the progress concerning the Quantum Monte Carlo algorithm for the $\Nf=2$ Hamiltonian.

\section{Hamiltonian formulation in the strong coupling limit for $N_f=2$}

\newcommand{\MU}{M_{\pi_U}}
\newcommand{\MD}{M_{\pi_D}}
\newcommand{\MP}{M_{\pi^+}}
\newcommand{\MM}{M_{\pi^-}}
\newcommand{\MPPUD}{M^{(2)}_{\pi^+\pi^-,UD}}
\newcommand{\MUDPP}{M^{(2)}_{UD,\pi^+\pi^-}}

\newcommand{\kU}{k_U}
\newcommand{\kD}{k_D}
\newcommand{\kP}{k_{\pi^+}}
\newcommand{\kM}{k_{\pi^-}}
\newcommand{\kPPUD}{k^{(2)}_{\pi^+\pi^-,UD}}
\newcommand{\kUDPP}{k^{(2)}_{UD,\pi^+\pi^-}}

While it is possible to derive a Hamiltonian formulation for gauge group $\SU(3)$ for any number of flavors, for definiteness we will here restrict to the formulation for $\Nf=2$ in the chiral limit. It should be noted that the number of hadronic states quickly grows with the number of flavors, the dimension $d$ of the local Hilbert space $\Hil$ is $d=6$ for $\Nf=1$, $d=92$ for $\Nf=2$ and $d=2074$ for $\Nf=3$.
The full Hilbert space has thus dimension $D=d^\Omega$ with $\Omega=N_s^3$ the spatial lattice volume. To refine the 92 states for $\Nf=2$ further in terms of baryon and isosopin number and meson occupation numbers, the 1-link integral is expressed via the following invariants \cite{Unger2021}:
\begin{align}
 \J(\QMat,\Qmd) &=\int\limits_{\SU(3)}\hspace{-2mm} dU e^{\tr[ U \Qmd+U^\dagger \QMat]}= \sum_{B=-2}^2 \sum_{n_1,n_2,n_3}
C_{B,n_1,n_2,n_3}
  \frac{E^B}{|B|!}\prod_{i=1}^3\frac{X_i^{n_i}}{n_i!},\quad E=
\begin{cases}
\det{\QMat}& B>0\\
1& B=0\\
\det{\Qmd}& B<0\\
\end{cases}
\nn
 (\QMat)_{ij}&=\bar{\chi}^\alpha_{i}(x) \chi^\alpha_i(y),\qquad (\QMat^\dagger)_{kl}=\chi^\beta_k(y)\bar{\chi}^\beta_l(x),\quad \Tr[(M_x M_y)^n] = (-1)^{n+1} \tr[(\QMat\Qmd)^n]\nn
X_1&=\Tr[M_x M_y]=\MU+\MD+\MP +\MM,\nn
 X_2&=X_1^2-D_2,\quad  X_3=X_1^3-2X_1 D_2,\nn
 D_2&=\det[M_x M_y]=\MU\MD +  \MP\MM - \MPPUD-\MUDPP\nn
 \det\QMat&= B_{uuu}+B_{uud}+B_{udd}+B_{ddd},\quad \det\Qmd =\bar{B}_{uuu}+\bar{B}_{uud}+\bar{B}_{udd}+\bar{B}_{ddd}
\label{JInt}
\end{align}
with $C_{B,n_1,n_2,n_3}$ combinatorial factors that are derived from \cite{Eriksson1981}, but expressed in a more suitable basis, in particular expressed in the $\Nf=2$ determinants $E$ and $D_2$. The $M_{\pi_i}$ are meson hoppings (with $\pi_1=\pi_U$, $\pi_2=\pi_D$, $\pi_3=\pi_{\pi^+}$, $\pi_4=\pi^-$) between nearest neighbor sites $\expval{x,y}$, the $B_{fgh}$ are baryons hopping from $x$ to $y$ and $\bar{B}_{fgh}$ anti-baryons hopping from $y$ to $x$. 
After Grassmann integration, negative weights occur within the invariant $X_2$, $X_3$ due to non-trivial Wick contractions from $D_2$. However, in the continuous time limit, only single meson exchange survives and in particular the two-meson hoppings $\MPPUD$, $\MUDPP$ can only appear in temporal direction. 
Without any resummations, there would be 287 possible states when considering all combinations of invariants from the $X_i$ and $E$ that survive after Grassmann integration. However, upon diagonalization of the transfer matrix from one set of states to another set, many states become resummed and only 92 distinct states survive. 
An example of such a tranfer matrix for $B=0$, $I=0$ is the square matrix which maps the states $\vec{e}_1=\ket{\MU\MD}$, $\vec{e}_2=\ket{\MP\MM}$, $\vec{e}_3=\ket{\MPPUD}$ and $\vec{e}_4=\ket{\MUDPP}$ onto each other:
\begin{align}
\Pi&=\left(
\begin{array}{llll}
\frac{9}{8} & -\frac{3}{8} & \frac{3\sqrt{3}}{8}  & -\frac{\sqrt{3}}{8} \\
-\frac{3}{8} & \frac{9}{8} & -\frac{\sqrt{3}}{3}  & \frac{3\sqrt{3}}{8} \\
-\frac{\sqrt{3}}{8} & \frac{3\sqrt{3}}{8} & -\frac{1}{8}  & \frac{3}{8} \\
-\frac{3\sqrt{3}}{8} & -\frac{\sqrt{3}}{8} & \frac{3}{8}  & -\frac{1}{8} \\
\end{array}
\right).
\end{align}
This matrix is a projector, has trace 2, and upon diagonlization, two linear combinations have eigenvalue $\lambda=0$ and can be disregarded, whereas the other two linear combinations have eigenvalue $\lambda=1$, which are the distinct states:
\begin{align}
\ket{\pi_1^2}&=\sqrt{3}\ket{\MU\MD}+\ket{\MUDPP}\nn
\ket{\pi_2^2}&=\sqrt{3}\ket{\MP\MM}+\ket{\MPPUD}
\label{piSq}
\end{align}
Also all other linear combinations that result from diagonalization have eigenvalues $\lambda=1$, few of them are two-fold-degenerated, and all result in positive weights. Hence, the sign problem is absent. 
In Tab.~\ref{HadronStatesNf2} the 92 quantum states are listed in terms of the the quantum numbers: baryon number $B$, isospin number $I$ and, number of mesons $\mathfrak{m}$. 
Those quantum numbers are not yet sufficient to distinguish all 92 states. 
\begin{table}
\begin{center}
\begin{tabular}{|r|r||c|c|c|c|c|c|c|c|c|c|c|c|c||c|}
\hline
$B$ & $I$\; & \multicolumn{13}{c||}{$\spin=\meson-\frac{3}{2}(2-|B|)$} & $\Sigma$ \\
\hline
 &  & $-3$ & $-\frac{5}{2}$ & $-2$ & $-\frac{3}{2}$ & $-1$ & $-\frac{1}{2}$ & $\, 0\,$ & $+\frac{1}{2}$ & $+1$ & $+\frac{3}{2}$ & $+2$ & $+\frac{5}{2}$ & $+3$ &  \\
\hline
\hline
 -2& 0 & & & & & & & 1 & & & & & & & 1 \\
\hline
 -1 & $-\frac{3}{2}$ &&&& 1 &&  1 && 1 && 1 &&&&  4 \\
 -1 & $-\frac{1}{2}$ &&&& 1 &&  2 && 2 && 1 &&&&  6 \\
 -1 & $+\frac{1}{2}$ &&&& 1 &&  2 && 2 && 1 &&&&  6 \\
 -1 & $+\frac{3}{2}$ &&&& 1 &&  1 && 1 && 1 &&&&  4 \\
\hline 
0 & -3 &  &&   &&& & 1 &&&&&&& 1\\ 
0 & -2 &  &&   && 1 && 2 && 1 && & && 4\\
0 & -1 &  && 1 && 2 && 4 && 2 &&  1 && & 10 \\
0 & 0  & 1&& 2 && 4 && 6 && 4 && 2 && 1 & 20 \\
0 & -1 &  && 1 && 2 && 4 && 2 &&  1 && & 10 \\
0 & -2 &  &&   && 1 && 2 && 1 && & && 4\\
0 & -3 &  &&   &&& & 1 &&&&&&& 1\\ 
\hline
 1 & $-\frac{3}{2}$ &&&& 1 &&  1 && 1 && 1 &&&&  4 \\
 1 & $-\frac{1}{2}$ &&&& 1 &&  2 && 2 && 1 &&&&  6 \\
 1 & $+\frac{1}{2}$ &&&& 1 &&  2 && 2 && 1 &&&&  6 \\
 1 & $+\frac{3}{2}$ &&&& 1 &&  1 && 1 && 1 &&&&  4 \\
\hline
 2 & 0 &&&&&&& 1 & & & & & & & 1 \\
\hline
\hline
$\Sigma$ &  &1 & 0 & 4 & 8 & 10 & 12 & 22 & 12 & 10 & 8 & 4 & 0 & 1 & 92\\    
\hline
\end{tabular}
\end{center}
\caption{
All 92 possible quantum states on a single site for the $\Nf=2$ Hamiltonian formulation with $\SU(3)$ gauge group. 
The number of states are given for the sectors specified baryon number $B$ and isospin number $I$, and symmetrized meson occupation number $\spin=\meson-\frac{\Nc}{2}(\Nf-|B|)$. Note the mesonic particle-hole symmetry  $\spin \leftrightarrow -\spin$ which corresponds to the shift symmetry by $\at$.
}
\label{HadronStatesNf2}
\end{table}

Since we are restricted to the chiral limit, a conservation law for each of the pion currents of $\mU$, $\mD$, $\mP$, $\mM$ holds. 
The role of spatial dimers at a bond location $\expval{x,y}$ is to transfer pion charge from one site $x$ to site $y$. Due to the even-odd ordering for staggered fermions, such dimers can be consistently oriented from an emission site $x$ to an absorption site $y$. 
As a consequence, if a occupation number $\meson_{\pi_i}(x)$ is raised/lowered by a spatial dimer, then at the site connected by the spatial meson hopping the meson occupation number $\meson_{\pi_i}(y)$ is lowered/raised.  With those interactions derived from a high temperature series, the resulting partition sum can be expressed in terms of a Hamiltonian that is composed of mesonic annihilation and creation operators $\hat{J}_Q^\pm$:
 \begin{align}
 Z_{\rm CT}(\bareT,\bareMu,\bareI,\Omega)&=\Tr_{\hadron^\Omega}\left[e^{(\hat{\calH}+\hat{\calN}_B\bareMu+\hat{\calN}_I\bareI)/\bareT}\right]\,\qquad \hadron \in \Hil\nn
 \hat{\calH}_I&=\frac{1}{2}\sum_{\langle\vec{x},\vec{y}\rangle}
 \sum_{\pii \in \{ \pi^+, \pi^-, \pi_U, \pi_D \}} \lr{
 {\hat{J}_{\pii,\vec{x}}}^+ {\hat{J}_{\pii,\vec{y}}}^- + {\hat{J}_{\pii,\vec{x}}}^- {\hat{J}_{\pii,\vec{y}}}^+%
 }\nn
 \hat{\calN}_B&={\rm diag}(-2,-1,\ldots 1,2),\qquad 
 \hat{\calN}_I={\rm diag}\lr{0,-\frac{3}{2},\ldots \frac{3}{2},0}
 \end{align}
 where the matrices per spatial site, ${\hat{J}_{\pii}}^+$, ${\hat{J}_{\pii}}^-$, $\hat{\calN}_B$ and $\hat{\calN}_I$ are $92\times 92$ - dimensional and the tensor product over all spatial sites $ \Omega$ is implied and $\Hil$ is the 92-dimensional local Hilbert space. 
For the transition $\hadron_1 \mapsto \hadron_2$, the matrix elements $\expval{\hadron_1|\hJ^{\pm}_{\pii} |\hadron_2}$ are determined from Grassmann integration and diagonalization, 
only those matrix elements are non-zero which are consistent with current conservation of all $\pii$.

Since meson occupation numbers are not just bounded from below, but also from above due to the Grassmannian nature of the underlying quarks, they fulfill an algebra that exhibits a particle-hole symmetry, the meson occupation numbers can be mapped onto a symmetrized occupation number: 
\begin{align}
\spin&=\meson-\frac{3}{2}(2-|B|)& \text{with} && \meson&=\sum_{i=1}^4 \meson_{\pi_i}.
\end{align}
On discrete lattices, particles are mapped $\spin\mapsto -\spin$  by a shift of $a_t$ due to the even-odd ordering of staggered fermions, but this relation also survives in the continuous time limit  $a_t\rightarrow 0$.\\

\begin{figure}[h!]
\centerline{\includegraphics[width=1.2\textwidth]{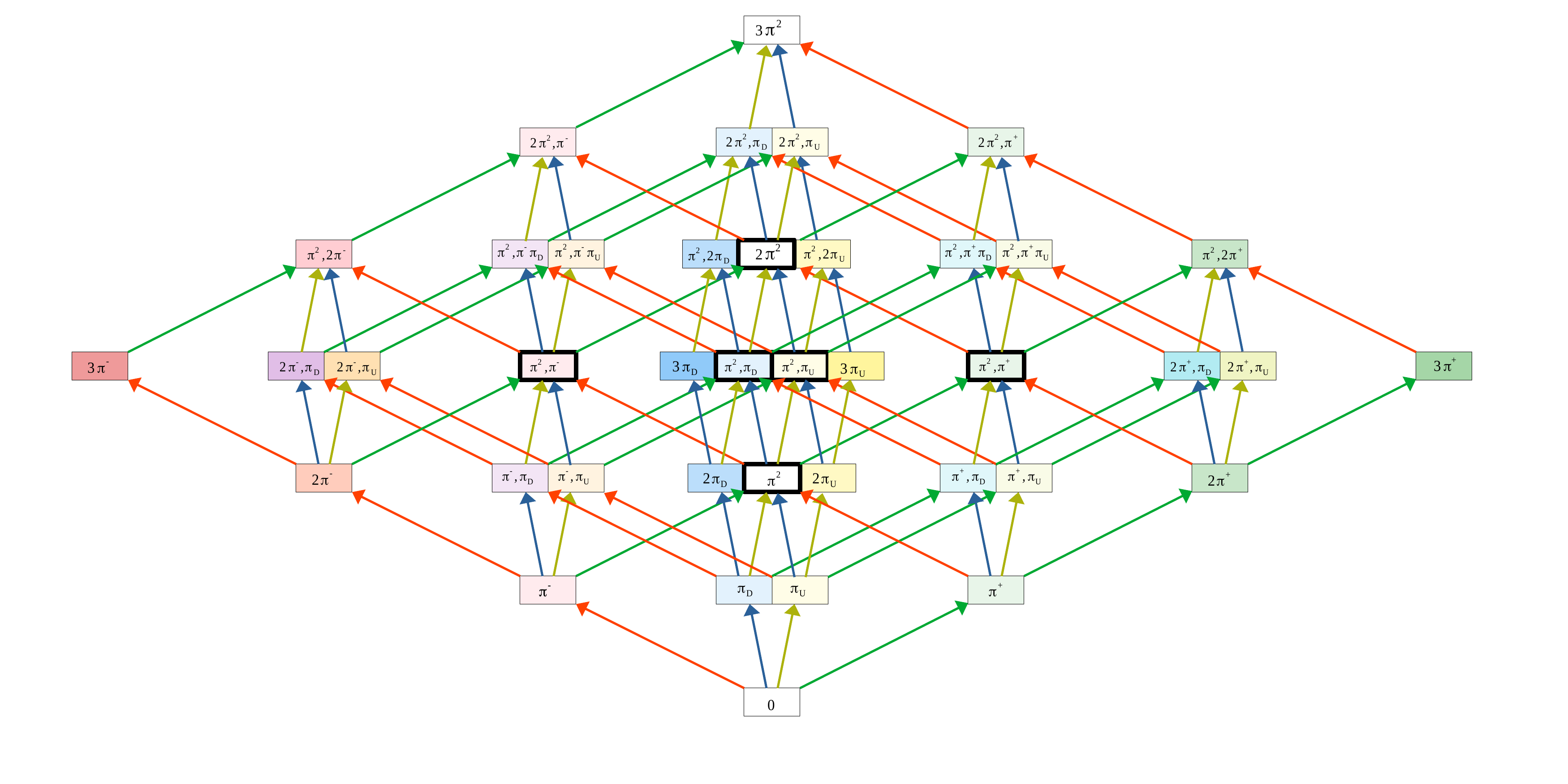}}
\vspace{-8mm}
\centerline{\includegraphics[width=1.2\textwidth]{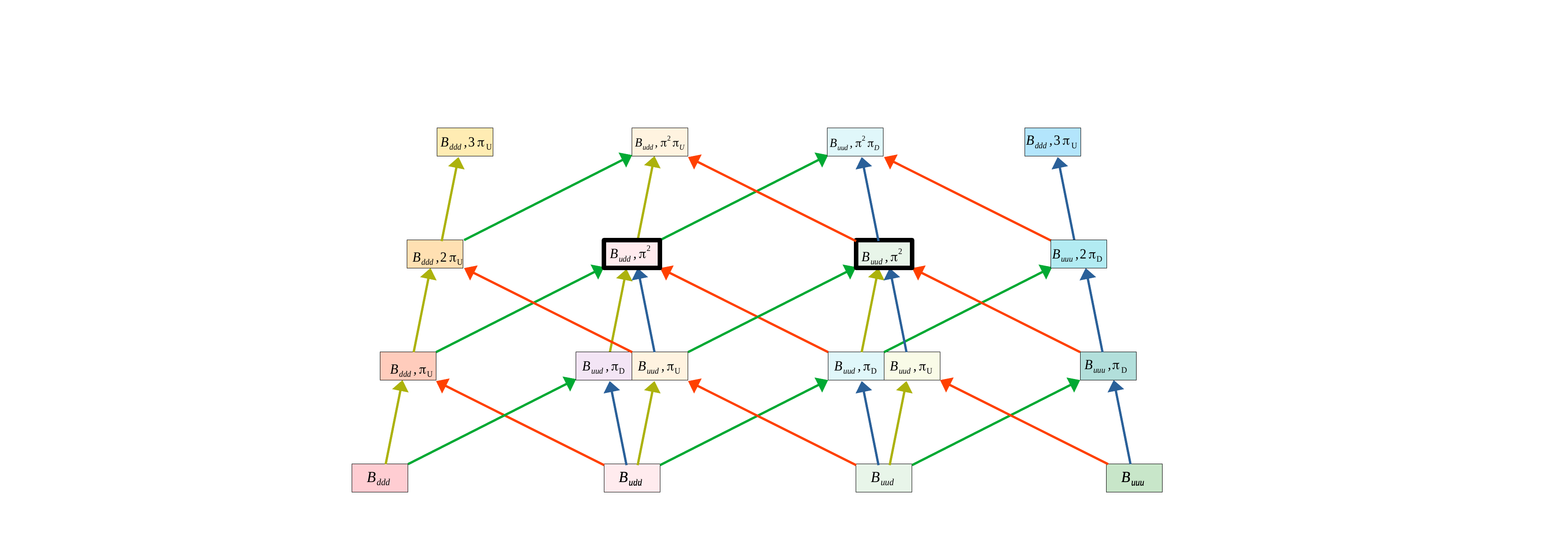}}
\caption{
\label{nf2classification}
Depiction of how ladder operators connect the various hadronic states, for $B=0$ (top) and $B=1$ (bottom).
The bold boxes are two-fold degenerated, see Eq.~(\ref{piSq}). The red arrows correspond to $\hJ^+_{\pi^{-}}$, the green arrows to 
$\hJ^+_{\pi^{+}}$, the blue arrows to $\hJ^+_{\pi_{D}}$ and the yellow arrows to $\hJ^+_{\pi_{U}}$. In the weight diagrams it can be seen that both the root system spanned by $\hJ^+_{\pi^+}$,$\hJ^+_{\pi^-}$ is orthogonal, and the root system spanned by 
$\hJ^+_{\pi_U}$,$J^+_{\pi_D}$ is orthogonal, both are direct product of two $\Nc+1$-dimensional representations, however, both system mix and cannot be treated independently. 
The horizontal axis label isospin, the vertical axis labels meson occupation numbers.
}
\end{figure}

The matrices $\hat{J}^\pm_{\pii}$ hence span a $\frac{3}{2}(2-|B|)$-dimensional representation of a Lie algebra, as illustrated in Fig.~\ref{nf2classification}. 
The arrows in different colors correspond to the raising ladder operators $\hat{J}^+_{\pii}$, each of the four colors generates a specific meson. The representation for $\hat{J}^\pm_{\pi_U}$ and $\hat{J}^\pm_{\pi_D}$ is a direct product representation, likewise 
$\hat{J}^\pm_{\pi^+}$ and $\hat{J}^\pm_{\pi^-}$, but both Lie algebras meet in various states, as they are not distinguished on the quark level, e.g.
 \begin{align*}
 \ket{\pi_+\pi_-}&=\ket{\pi_U\pi_D},&
 \ket{B_{uuu}\pi_D}&=\ket{B_{uud}\pi_-},&
 \ket{B_{uuu} B_{ddd}}&=\ket{B_{uud} B_{udd}}.
 \end{align*}
Those states in Fig.~\ref{nf2classification} that are twofold degenerated as for $\ket{\pi_1^2}$, $\ket{\pi_2^2}$ in Eq.~(\ref{piSq}) are highlighted in bold: 5 such states for $B=0$ and 2 states degenerate for $B=1$ and also for $B=-1$.
We label all 92 hadronic states of the local Hilbert space by their quark content in lexicographical order:
first ordered by $B$, $I$ and $\meson$ and then by the sequence of occupations in $\bar{u}$, $u$, $\bar{d}$, $d$.
However, the quark content is not sufficient to distinguish those 9 states that are two-fold degenerate: here we introduce an additional index $i\in\{0,1\}$ that is required by the QMC algorithm as explained in the next section. 

The ladder operators fulfill the following identities:
\begin{align}
[\hat{J}^+_{\pii},\hat{J}^-_{\pii}]&=\left(
\begin{array}{cccc}
-1 & 0 & 0 & 0  \\
0 & -1/3 & 0 & 0  \\
0 & 0 & 1/3 & 0 \\
0 & 0 & 0 & 1  \\
\end{array}\right)=\frac{2}{\Nc}\hat{J}^{(3)}_{\pii}, &
[\hat{J}^+_{\pii},\hat{J}^-_{\pi_j}]&=0\quad \text{for}\quad  i\neq j.
\end{align}, 

Examples of matrix elements of $\hJ^+_{\pii}$ (with $\pi_1=\pi_U$, $\pi_2=\pi_D$, $\pi_3=\pi^+$ and  $\pi_4=\pi^-$) are:
\begin{align}
\braOket{\pi_i}{\hJ^+_{\pii}}{0}&=1, \quad  \braOket{2\pi_i}{\hJ^+_\pii}{\pi_i}=\frac{2}{\sqrt{3}}, \quad \braOket{3\pi_i}{\hJ^+_{\pii}}{2\pi_i}=1 \quad \text{for}\quad i\in\{1,2,3,4\}\nn
\braOket{\pi_i\pi_j}{\hJ^+_{Q_j}}{\pi_i}&=1\quad \text{for}\quad (i,j)\in\{(1,3),(1,4),(2,3),(2,4)\}\nn
\braOket{\pi_1^2}{\hJ^+_{\mD}}{\mU}&=\braOket{\pi_1^2}{J^+_{\mU}}{\mD}=\frac{\sqrt{6}}{4},\quad 
\braOket{\pi_2^2}{J^+_{\mD}}{\mU}=\braOket{\pi_2^2}{J^+_{\mU}}{\mD}=-\frac{\sqrt{6}}{12},\nn
\braOket{\pi_2^2}{\hJ^+_{\pi^-}}{\mP}&=\braOket{\pi_2^2}{\hJ^+_{\pi^+}}{\mM}=\frac{\sqrt{6}}{4}\quad 
\braOket{\pi_1^2}{\hJ^+_{\pi^-}}{\mP}=\braOket{\pi_1^2}{\hJ^+_{\pi^+}}{\mM}=-\frac{\sqrt{6}}{12}
\end{align}
Although the matrix elements involving cross-terms are negative, since $\pi_1^2$ and $\pi_2^2$ are not distinguished on the quark level,
any other linear combination will also work. 
With the symmetric and anti-symmetric linear combination
\begin{align}
\ket{\pi_0^2}=\frac{1}{2}\lr{\ket{\pi_1^2}+\ket{\pi_2^2}}, \qquad \ket{\bar{\pi_0}^2}=\frac{1}{2}\lr{\ket{\pi_1^2}-\ket{\pi_2^2}}
\end{align} 
we find
\begin{align}
\braOket{\pi_0^2}{\hJ^+_{\mD}}{\mU}&=
\braOket{\pi_0^2}{\hJ^+_{\mU}}{\mD}
=\braOket{\pi_0^2}{\hJ^+_{\mM}}{\mP}
=\braOket{\pi_0^2}{\hJ^+_{\mM}}{\mM}
=\frac{1}{2\sqrt{6}},\nn
\braOket{\bar{\pi}_0^2}{\hJ^+_{\mD}}{\mU}&=
\braOket{\bar{\pi}_0^2}{\hJ^+_{\mU}}{\mD}
=\braOket{\bar{\pi}_0^2}{\hJ^+_{\mM}}{\mP}
=\braOket{\bar{\pi}_0^2}{\hJ^+_{\mM}}{\mM}
=\frac{1}{\sqrt{6}}
\end{align}
All the other matrix elements can be consistently combined to result in only positive values.

\section{Dependence on the chemical potential}

\begin{figure}
\includegraphics[width=0.49\textwidth]{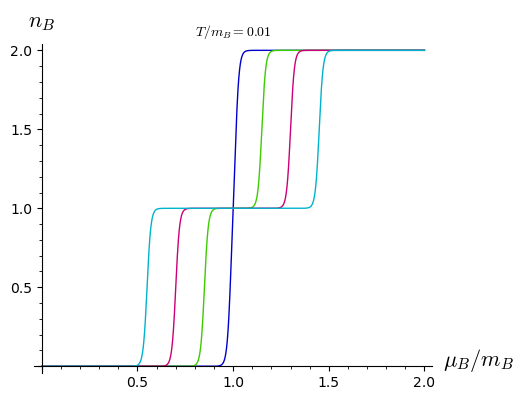}
\includegraphics[width=0.49\textwidth]{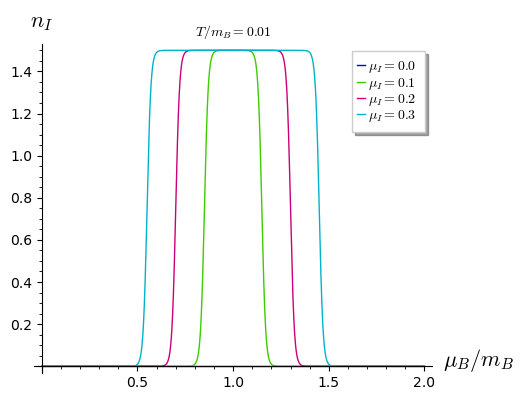}
\caption{
\label{nBnI}
Result on the baryon density (left) and isospin density (right) at non-zero isospin chemical potential in the static limit (corresponding to the high temperature limit at strong coupling). 
}
\end{figure}

In the static limit, which corresponds in our setup to the high temperature limit where pion exchange is absent, we have $
Z=Z_1^V$ with $Z_1$ is the 1-dim.~QCD partition function.
All 92 states $\hadron\in \Hil$ contribute with a weight that depends on the baryon and isospin chemical potential. Based on the Conrey-Farmer-Zirnbauer formula \cite{Ravagli2007} we have derived $Z_1$
or degenerate quark mass $m \equiv m_u=m_d$, with $\mu_c=\mu_c(m)$ the effective mass as a function of the bare mass:
\begin{align}
Z1\lr{\frac{\mu_B}{T},\frac{\mu_I}{T},\frac{\mu_c}{T}}&=2\cosh\frac{3\mu_I}{T}+4\lr{\cosh\frac{\mu_c}{T}}^2\lr{3+2\cosh\frac{4\mu_c}{T}+2\cosh\frac{2\mu_I}{T}}\nn
&+4\cosh\frac{\mu_I}{T}\lr{2+2\cosh\frac{2\mu_c}{T}+\cosh\frac{4\mu_c}{T}}\nn
&+8\cosh\frac{\mu_B}{T}\lr{2\cosh\frac{\frac{3}{2}\mu_I}{T}\cosh\frac{\mu_c}{T}+
\cosh\frac{\frac{1}{2}\mu_I}{T}
\lr{2\cosh\frac{2\mu_c}{T}+1}}\nn
&+2\cosh\frac{2\mu_B}{T}
\label{Zirnbauer}
\end{align}
The Quantum Hamiltonian at finite quark mass still only contain 92 hadronic states per site, but a set of annihilation/creation operators on a single site need to be included, which we will discuss in a forthcoming publication.
In Fig.~\ref{nBnI} the baryon density and isospin density as obtained by taking derivatives from Eq.~({\ref{Zirnbauer}}) is show for various isospin chemical potentials at fixed baryon chemical potential and temperature. 
We find that for $\mu_I>0$ the baryon density $n_B$ signals two transitions, the first taking place when the isospin density jumps to its maximal value $n_I=3/2$, the second transition taking place when the isospin density vanishes again, which is due to Pauli saturation. A non-zero isospin density does only admit a single baryon, but not $n_B=2$. This behaviour is consistent with $\Nf=2$ meanfield theory for staggered fermions \cite{Nishida2003}. Here it was found that at non-zero isospin density, two critical end-points exist, at the first transition the condensate $\sigma_u$ vanishes in the second transition also  $\sigma_d$ vanishes.
We aim to confirm this scenario with Monte Carlo simulations. 

\section{Setup of the Quantum Monte Carlo Simulation}

The $\Nf=2$ QMC algorithm is an extension of the $\Nf=1$ QMC and is also realized as a continuous time version of the worm algorithm for strong coupling LQCD \cite{Klegrewe2020}. We will focus here mainly on the modification required for $\Nf=2$:
\begin{enumerate}
 \item For the initial configurations, at every spatial site one of the 92 states is selected with a weight according to the values of the chemical potentials 
 $\mu_B$, $\mu_I$. 
 \item Prior to the worm updates, a specific meson $\pii$ from the four possible states $\{\mU,\mD,\mP,\mM\}$ has to be chosen randomly, and both $\hJ^+_{\pii}$, $\hJ^-_\pii$ will be fixed during worm evolution until the worm closes. 
 \item The continuous time worm update: 
 \begin{enumerate}
 \item The move update: choose a new admissible site $(x_T,t_T)$ for worm head and tail to start the Poisson process.
 \item The shift update: during the Poisson process the worm head moves continuously in Euclidean time (positive or negative direction with the possibility to wrap around due the periodic boundary) until it emits or absorbs a spatial pion of charge $\pi_i$. 
  The emission (``decay``) probabilities $\exp(-\lambda_\pii(t) \Delta_t)$ at some time $t$ from site $x$ to $y$ is given by
 \begin{align}
 \lambda_\pii(t)&\sim \frac{\braOket{\hadron^{\rm in}_{x,t}}{\hJ^+_{\pii,x}}{\hadron^{\rm out}_{x,t}}\braOket{\hadron^{\rm in}_{y,t}}{\hJ^-_{\pii,y}}{\hadron^{\rm out}_{y,t}}}{T},
 \end{align}
 with $T$ the temperature. The lower the temperature, the more interactions are generated. The decay constant $\lambda(t)$, in contrast to $\Nf=1$, depends on time: the number of admissible neighbors of $x$ from which the site $y$ is chosen depends on the hadronic state $\hadron_y(t)$, in particular on the flavor content during time $\Delta_t$ as it may be blocked for pion exchange.
 \item The worm recombines when the worm head returns to $(x_T,t_T)$.
 \end{enumerate}
 \item A static update is probed for all sites to which no spatial pion is attached: again, for those sites a new of the 92 hadronic states is chosen, which may change the baryon number (that cannot be changed during worm evolution as baryons are static in the continuous time limit).
 \item The next worm update is proposed, starting at 2.~and repeated until the desired statistics is reached.
 \end{enumerate}
The baryon and isospin density can be measured on each configuration after worm update, by averaging over time slices. Also the chiral and pion susceptibilities can be obtained from the integrated 2-point correlation functions measured during worm evolution as so-called improved estimators. First simulations in small volumes, by scanning in baryon chemical potential at fixed isospin chemical potential and have indeed found a plateau between the two transitions that increases with increasing isospin density. As we are still preparing simulations in larger volumes and for lower temperatures, we will present data on the $\Nf=2$ nuclear transition in a forthcoming publication.

\section{Summary and Outlook}


We have presented an extension to the Hamiltonian formulation of strong coupling lattice QCD from $\Nf=1$ to $\Nf=2$ and gave a detailed account of the way the hadronic states are used in a QMC algorithm. The simulations are sign-problem free. 
We are still in the process to map out the enlarged phase diagram in the $\mu_B-\mu_I-T$-space and will present results on the nuclear and chiral transition and pion condensation in a forthcoming publication. 

We plan to extend this framework in two directions: (1) by including modifications due to finite quark masses, and (2) by including the gauge corrections to the strong coupling limit. 
Whereas (1) will not alter the number of hadronic states, but will add new interactions between the hadronic states, (2) will also add new quantum states which are not purely hadronic, but involve combinations of quarks and gluons as color singlets on which creation and annihilation operators act upon. It is not yet guaranteed that these extensions are sign-problem free, but it is in any case much milder than on a lattice with discrete time. 

This work was supported by the Deutsche Forschungsgemeinschaft (DFG, German Research Foundation) – project number 315477589 – TRR 211.

\end{document}